\documentclass[fleqn,10pt]{wlscirep}
\title{Testing Modeling Assumptions in the West Africa Ebola Outbreak}

\author[1,*]{Keith Burghardt}
\author[2]{Christopher Verzijl}
\author[3,4]{Junming Huang}
\author[5]{Matthew Ingram}
\author[6]{Binyang Song}
\author[7]{Marie-Pierre Hasne}

\affil[1]{Department of Political Science, University of California at Davis, Davis, 95616, USA}
\affil[2]{ABN AMRO Bank N.V., Amsterdam, 1082 PP, Netherlands}
\affil[3]{Comple$\chi$ Lab, Web Sciences Center, University of Electronic Science and Technology of China, 611731, P. R. China}
\affil[4]{Center for Complex Network Research, Department of Physics, Northeastern University, Boston, 02115, USA}
\affil[5]{Department of Political Science and Center for Social and Demographic Analysis, University at Albany, State University of New York, Albany, 12203, USA}
\affil[6]{SUTD-MIT International Design Centre \& Engineering Product Development Pillar, Singapore University of Technology and Design, Singapore, 487372, Singapore}
\affil[7]{Department of Biochemistry and Molecular Biology, Oregon Health \& Science University, Portland, 97239, USA}

\affil[*]{kaburghardt@ucdavis.edu}

\begin{abstract}
The Ebola virus in West Africa has infected almost 30,000 and killed over 11,000 people. Recent models of Ebola Virus Disease (EVD) have often made assumptions about how the disease spreads, such as uniform transmissibility and homogeneous mixing within a population. In this paper, we test whether these assumptions are necessarily correct, and offer simple solutions that may improve disease model accuracy. First, we use data and models of West African migration to show that EVD does not homogeneously mix, but spreads in a predictable manner. Next, we estimate the initial growth rate of EVD within country administrative divisions and find that it significantly decreases with population density. Finally, we test whether EVD strains have uniform transmissibility through a novel statistical test, and find that certain strains appear more often than expected by chance.
\end{abstract}
\begin{document}

\flushbottom
\maketitle
\thispagestyle{empty}

%\noindent Please note: Abbreviations should be introduced at the first mention in the main text – no abbreviations lists. Suggested structure of main text (not enforced) is provided below.

\section*{Introduction}

Since 2013, the Ebola virus outbreak in West Africa has become the largest such outbreak known. The epidemic first emerged in December 2013 in southern Guinea, but as of 11 May 2016, there have been about 28,600 cases of EVD in Guinea, Liberia, and Sierra Leone, and isolated cases in Italy, Mali, Nigeria, Senegal, Spain, the United Kingdom, and the United States. 11,300 of these cases were fatal \cite{WHOEbola} and, as high as these numbers are, they may be under-estimates due to the poor quality of current data \cite{Spatiotemporal}. The goal of this paper is to better understand the spread of EVD, and test the assumptions of leading EVD models.

Individuals have often been assumed to homogenously mix with each other in many recent EVD models \cite{EbolaMixingModel1,EbolaMixingModel2,EbolaMixingModel3,EbolaMixingModel4,EbolaMixingModelReview}, but we show that, by applying recent work on the migration of diseases \cite{ContagionGeometry}, homogeneous mixing is an especially poor approximation for EVD. We find that human migration patterns help predict where and when EVD originated and will appear, which would not be possible with a homogeneous mixing assumption. We also find evidence that the spread of EVD is much slower than other recent diseases, such as H1N1 and SARS \cite{ContagionGeometry}, which may have helped health workers control the disease.

Furthermore, against our expectations, we find that the initial growth rate of EVD can decrease significantly with population density, possibly because higher population density areas are correlated with other attributes, such as better healthcare. This compares to previous work where exponential and sub-exponential growth rates were found in many diseases, including the most recent EVD epidemic \cite{Chowell15,Chowell16}, where variations in the growth rate of diseases were found, but mechanistic explanations were not explored. A previous model \cite{ScalingHumanInteractions}, in comparison, found that higher population cities should contribute to a faster rate of disease spread, although we are not aware of previous research on disease spread and population density. Our work suggest that location-specific initial growth rates better model EVD, although the underlying reason for this heterogeneity should be a topic of future research. 

Finally, we create novel metrics for the relative transmissibility of EVD strains, which are robust to sparse sampling. These metrics add to previous work on EVD in Sierra Leone \cite{EbolaCompetingClaudes}, and provide a novel understanding of EVD strains in Guinea. We find that the relative transmissibility of strains, as measured from these metrics, is not uniform; therefore, treating EVD as a single disease may be inappropriate \cite{EbolaMixingModel1,EbolaMixingModel2,EbolaMixingModel3,EbolaMixingModel4,EbolaMixingModelReview}.

These results, when taken together, suggest unexpectedly simple ways to improve EVD modeling. In the Discussion section, we will explain how a meta-population model can potentially aid in our understanding of disease spread and growth. Furthermore, incorporating disease strain dynamics into this model could help us better predict which strains will become dominant in the future, which may improve vaccination strategies.

%%%%%%%%%%%%%%%%%%%%%%%%%%%%%%%%
\section*{Results}
Models of the West Africa Ebola outbreak have often assumed that the disease spreads via homogeneous mixing \cite{EbolaMixingModel1,EbolaMixingModel2,EbolaMixingModel3,EbolaMixingModel4,EbolaMixingModelReview}. We find, however, that this assumption may not accurately model EVD when the disease first arrives in a given area. We will first discuss how the arrival time of EVD within a country or administrative area follows a predictable pattern due to the underlying migration model, in contrast to the mixing hypothesis. Next, we model the cumulative number of individuals infected in administrative divisions at the first or second level in Guinea, Liberia, and Sierra Leone to estimate the initial growth rate of EVD. We find this growth rate varies significantly, and appears to decrease with the population density within the administrative division. Finally, we introduce models of how EVD disease strains spread to rule out uniform strain transmissibility.

\subsection*{How Does Ebola Spread?}

\begin{figure}[hbt]
\centering
 \includegraphics[width=0.8 \columnwidth]{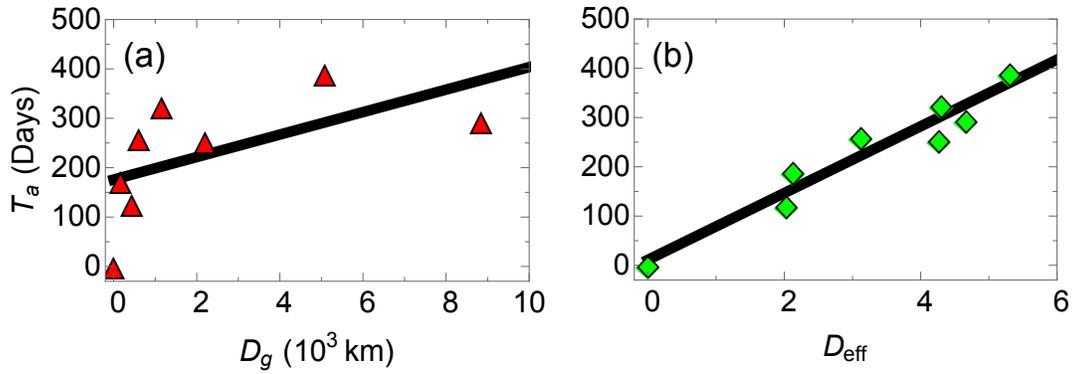}
\caption{\label{fig:DeffCountry}The arrival time, $T_a$, of EVD in a country versus (a) the great-arc length distance, $D_\text{g}$ and (b) the migration-based effective distance, $D_\text{eff}$ from the disease's point of origin (Guinea). Error is smaller than marker size. The migration network used to construct the effective distance comes from census microdata \cite{EbolaMigration}. The linear relationship between arrival time and effective distance suggests that there is a constant effective velocity of disease spread, in agreement with previous work on other diseases \cite{ContagionGeometry}.}
\end{figure}

Homogeneous mixing models assume that healthy individuals can get sick regardless of where they are, even when they are hundreds of miles from the origin of the infection. If this is true, then the disease should be quickly seen in all susceptible areas almost simultaneously. Although this approximation may be reasonable at short distances, there has to be a lengthscale when this would break down because, in the years since Ebola first emerged, no more than a handful of countries have become infected (Fig. \ref{fig:DeffCountry}). 

Alternatively, one might assume that EVD spreads spatially. There is a significant positive correlation (Spearman $\rho = 0.26$, $p<0.05$, $n = 56$ at the spatial resolution of administrative divisions; $0.81$, $p<0.05$, $n = 8$ at the country resolution) between the arrival time of EVD in administrative divisions at the first or second level and the distance from the outbreak origin, Gu\'eck\'edou, Guinea, to division centroids. Furthermore, this assumption has been applied successfully to model EVD in Liberia \cite{Spatiotemporal}. 

We find, however, that migration in West Africa is more complex than either of these assumptions \cite{EbolaMigration}. Intuitively, diseases should spread more quickly between administrative divisions or countries with significant travel between them than between isolated areas. Therefore, it is reasonable to rescale distances such that areas with stronger travel ties are considered ``closer'' than areas without these ties, following work by Brockman and Helbing \cite{ContagionGeometry}. We find that rescaling distances using migration patterns helps us (1) better understand how quickly EVD spreads, and (2) estimate where the outbreak started. 

\begin{figure}[hbt]
\centering
 \includegraphics[width=0.8 \columnwidth]{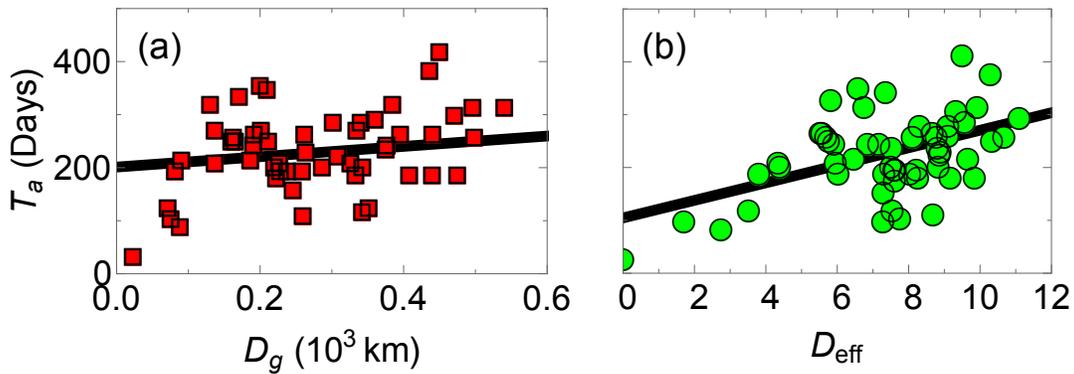}
\caption{\label{fig:DeffDistrict}The arrival time, $T_a$, of patients with EVD in administration divisions at the first or second level within Guinea, Mali, Liberia, Sierra Leone, or Nigeria, versus (a) the great-arc length distance, $D_\text{g}$, and (b) the migration-based effective distance, $D_\text{eff}$, from the disease's point of origin (Gu\'eck\'edou, Guinea). Error is smaller than marker size. The migration network used to construct the effective distance comes from a radiation migration model \cite{MigrationModel} (similar results were found using the gravity migration models in \cite{EbolaMigration}, see Supplementary Fig. S1 online).}
\end{figure}

To our surprise, we find that the correlation between the arrival time and effective distance for EVD (Spearman $\rho = 0.95$, $p<10^{-3}$, $n = 8$) was consistently higher than the correlation between the arrival time and geodesic distance (see Figs.~\ref{fig:DeffCountry}, \ref{fig:DeffDistrict}, and Supplementary Fig. S1 online). A high Pearson correlation ($0.96$, $p<10^{-3}$), and agreement with the normality assumption ($p > 0.05$, using the Kolmogorov-Smirnov normality test), also suggests a linear relationship between the arrival time and effective distance. Taking the slope of the plot gives us an effective velocity of spread, which we find to be $0.015 \text{days}^{-1}$, which is much slower than for previous diseases ($\approx0.1 \;\text{days}^{-1}$) \cite{ContagionGeometry}. 
An intuitive explanation for our finding is that lower overall migration in West Africa reduces the speed at which EVD spreads compared to other diseases. 

\begin{figure}[hbt]
\centering
 \includegraphics[width=0.4 \columnwidth]{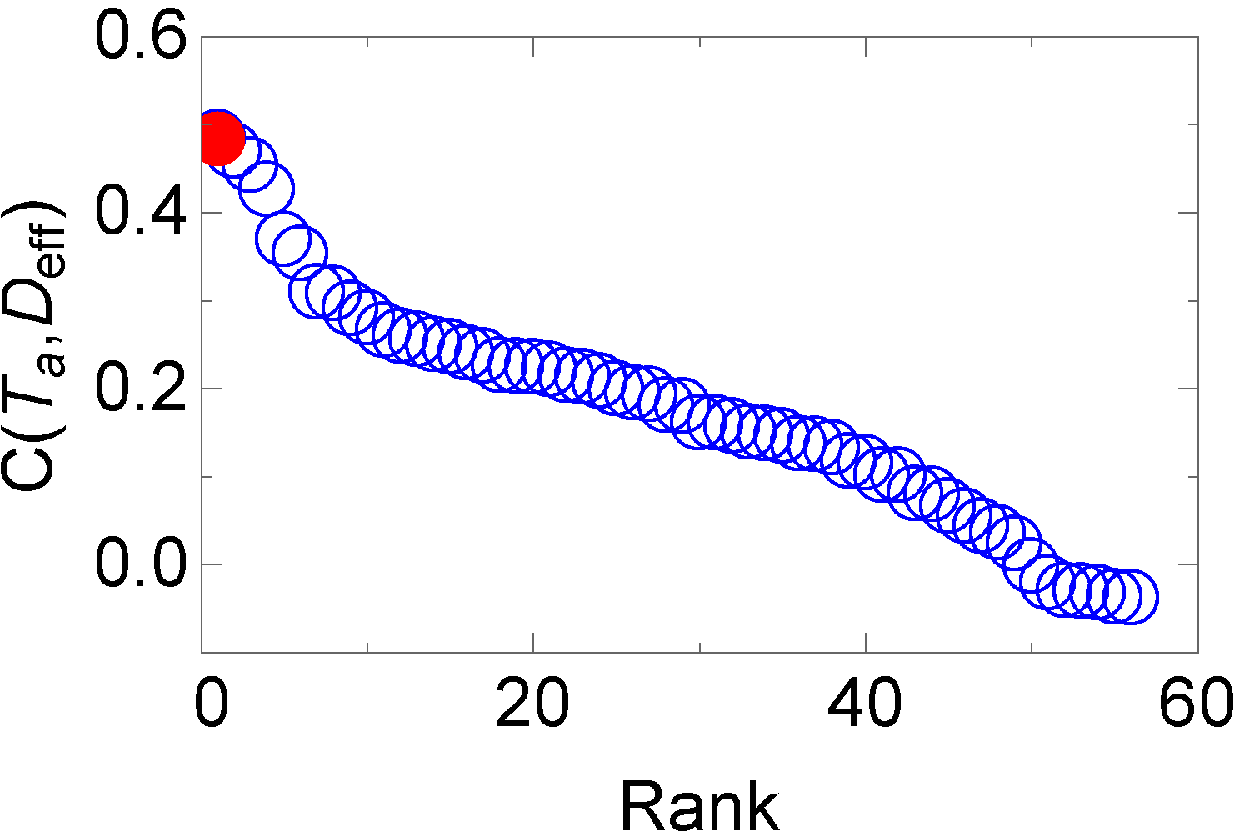}
\caption{\label{fig:DistrictRank}
 The (ranked) correlations between arrival time, $T_a$, and migration-based effective distance, $D_{eff}$, between administrative divisions, calculated via Eq.~5 in \cite{ContagionGeometry}. The migration network used to construct the effective distance comes from a radiation migration model \cite{MigrationModel} (similar results were found using the gravity migration models in \cite{EbolaMigration}, see Supplementary Fig. S2 online). In red is the true origin, which is found to have the highest correlation.
}
\end{figure}

The lower correlations at the first or second administrative level are due in part to the fact that we use migration models to determine the effective distance, and the disease spread for several months before it was detected \cite{EbolaMixingModelReview}. Despite the lower quality data, however, we can still use it to determine the most likely origin of EVD, which we compare to the known origin, Gu\'eck\'edou, Guinea.  Quickly finding where a disease originated is important to help understand what caused it (e.g., what was the vector), and to predict where and when it will arrive, which can allow health workers to prepare \cite{ContagionGeometry}. Previously, it was found that the correlation between the arrival time and the effective distances from the disease origin is higher than correlations from areas where the disease did not originate \cite{ContagionGeometry}. We therefore ranked the correlation between the arrival times and effective distances from all the administrative divisions where Ebola was found between 2013 and 2016 (Figs.~\ref{fig:DistrictRank} \& Supplementary Fig. S2 online), and found the true origin has the highest correlation. To determine the statistical significance of the ranking, we bootstrapped residuals of the linear regression between $D_{eff}$ and $T_a$, with Gu\'eck\'edou as the origin, to accurately bootstrap arrival times. This is equivalent to simulating what would happen if we wound back the clock and restarted the infection from Gu\'eck\'edou. The true origin has the highest correlation $45\%\pm 1.6\%$ of the time for the radiation model, and $44\%-70\%$ of the time for gravity models (Supplementary Fig. S2 online), therefore the true origin does not always have the highest correlation, but it often does, and can be a good first guess when determining the origin.

In conclusion, we find strong evidence that a migration network can elucidate how quickly Ebola spreads, when and where it will arrive, and where the infection began even with limited data. Furthermore, we see that alternative hypotheses for how EVD spreads, such as homogeneous mixing and nearest neighbor interactions, provide quantitatively poorer agreement with data.

\subsection*{The Growth of EVD Across Administrative Divisions}

\begin{figure}[hbt]
\centering
 \includegraphics[width=0.6\columnwidth]{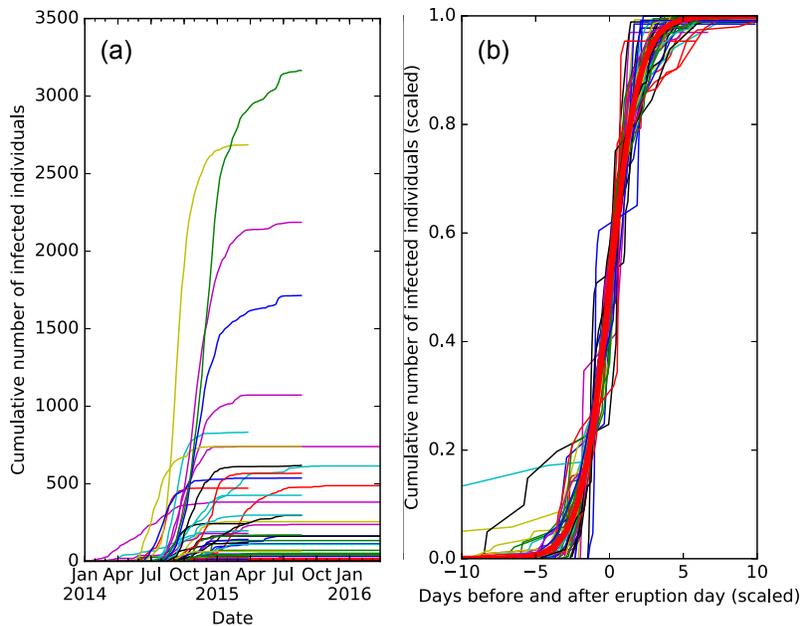}%raw-and-scaled-cumulative-infection.eps}
\caption{\label{fig:original-scaled-infection-curve-over-time} The cumulative number of infected individuals over time within administration divisions at the first or second level in Guinea, Sierra Leone and Liberia, from the patient database dataset \cite{WHOEbola}. It is clear that the rate and size of infections are heterogeneous, but when we fit the data to logistic functions, and renormalize the coefficients to $p_n\rightarrow 1$, $\tilde{t} = (t-t_0) q_n$ in (b), we see that the distributions collapse. A logistic function is plotted in red.}
\end{figure}

In this section, we show that the cumulative number of Ebola cases within administrative divisions at the first or second level is well-approximated by a logistic function (Fig.~\ref{fig:original-scaled-infection-curve-over-time}a). We use this finding to estimate the initial growth rate of EVD for each infected administrative division, where, to our surprise, we find that the initial growth rate decreases with population density. We emphasize that variations in EVD growth rates have been seen before \cite{Chowell15,Chowell16}, but mechanisms that might drive this behavior were not proposed. The logistic function has been used to model initial stages of EVD \cite{Chowell14}, and is equivalent to the Susceptible-Infectious (SI) model when 100\% of individuals are initially susceptible \cite{Bailey75}. To fit the SI model to our data, however, we would have to make a simplistic assumption that only a fraction $p_n$ of individuals are susceptible, in order to explain why a small fraction of the population is ever infected. We do not claim this describes the actual dynamics of EVD, although we will explain later why the cumulative number of cases should approximately follow this distribution. For an administrative division $n$, the cumulative number of infected individuals over time, $t$, by the logistic growth function is simply: 

\begin{equation}
\label{LogisticRegression}
i_n(t)=\frac{p_n}{1+e^{-q_n(t-{t_0}_n)}}.
\end{equation}

The initial growth rate is $q_n$, while $t_0$ is the time when the cumulative infection increases the most.  The dynamics are highly heterogeneous (Fig.~\ref{fig:original-scaled-infection-curve-over-time}a), therefore, it would seem un-intuitive for a single function to fit all the data. However, after fitting each cumulative distribution to the logistic model, and rescaling the variables: $p_n\rightarrow 1$, $\tilde{t} = (t-t_0) q_n$, we find that the distributions collapse (Fig.~\ref{fig:original-scaled-infection-curve-over-time}b) to a good approximation.

\begin{figure}[htb]
\centering
\includegraphics[width=0.5 \columnwidth]{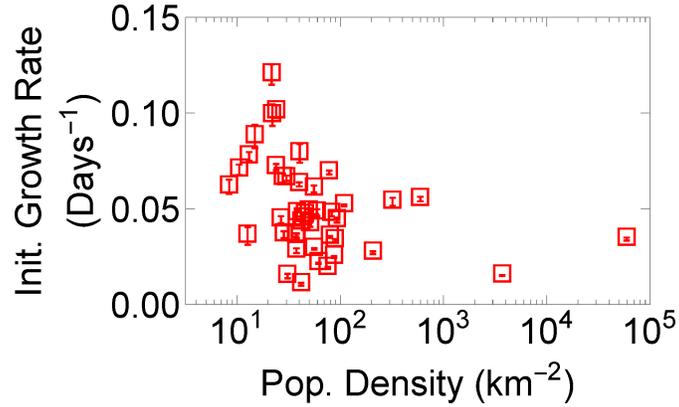}%EbolaGrowthVsPopDensity_SI_PSD.eps}
\caption{\label{fig:ExpVsPopDen}The initial growth rate, $q_n$, versus population density for the PSD dataset with more than 20 infections (not shown is the outlier Kissidougou, Guinea, with a population density of 34 individuals per kilometer, and a growth rate of $0.41 \text{days}^{-1}$), where error bars are standard deviations. We find that the initial growth rate drops significantly with population density (Spearman $\rho = -0.48$, $p< 10^{-3}$, $n = 44$).}
\end{figure}

Because small variations in $q_n$ very quickly become substantial variations in the infection size later on,  we want to understand how $q_n$ varies across administrative divisions. When plotting $q_n$ versus population density (Fig.~\ref{fig:ExpVsPopDen}) with more than 20 infections total, we notice that, although $q_n$ is $\approx0.1~\text{days}^{-1}$, which is similar to a previous EVD outbreak \cite{EbolaGrowthRate}, $q_n$ varies significantly across administrative divisions. This result contrasts with many previous models that assume a global parameter can describe the growth of Ebola \cite{EbolaMixingModel1,EbolaMixingModel2,EbolaMixingModel3,EbolaMixingModel4,EbolaMixingModelReview}. Furthermore, we find that $q_n$ decreases significantly with population density (Spearman $\rho = -0.48$, $p< 10^{-3}$, $n = 44$), plausibly because better healthcare may exist in higher density areas. This contrasts with a previous model, which predicted a positive scaling relation between the mean growth rate and population of cities \cite{ScalingHumanInteractions}, although we are not aware of research that explored how disease spread is affected by population density. Therefore, not only is the growth rate of Ebola unexpectedly heterogeneous, but the dependence on population density may help us understand why this is the case.

\paragraph{What Makes the Data Collapse?}

%The next step to a less minimal but more realistic epidemic model is the extension to an 
In the SI model, $100\%$ of individuals are eventually infected, therefore, to find agreement with data, our model has to assume a small proportion of individuals in each division are susceptible to the disease. This seems implausible; more likely, all individuals are susceptible and, as they become aware of an infection, they reduce their interactions or otherwise reduce the overall disease transmissibility. To demonstrate this hypothesis, we created a more realistic, although still simplistic, disease model, in which susceptible (s) individuals can become infected (i), but then recover or are removed \cite{SIR}. In our data, all individuals who recovered or died were assumed to be ``removed". The SIR model, like the SI model, significantly overshoots the cumulative number of cases in the absence of intervention. We counter-balance this effect with an exponentially decreasing disease transmissibility as a result of public health interventions \cite{SDIR1,SDIR2}. We call this model the Susceptible - Decreasingly Infectious - Recovered (SDIR) model. 

\begin{figure}[hbt]
\centering
\includegraphics[width=0.8\columnwidth]{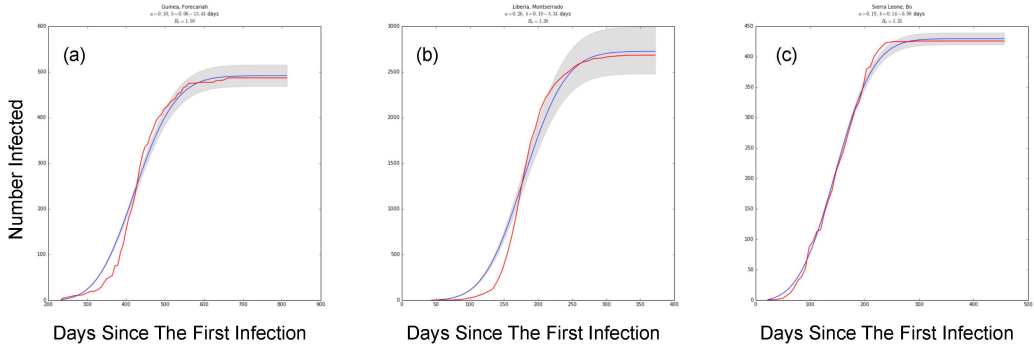}%Fit_Sierra_Leone_1_Western_Area_Urban.eps}
\caption{\label{fig:all-country-fits} SDIR fits to empirical data (fits for all infected administrative divisions can be seen in Supplementary Figs. S3, S4, \& S5 online). The cumulative number of infected individuals in (a) Forecariah, Guinea; (b) Montserrado, Liberia; and (c) Bo, Sierra Leone, where the blue line is the best fit and the shaded area represents standard errors. While some fits are poorer than others, we generally capture the qualitative structure of the empirical data, including the cumulative number of infected. This model may therefore begin to explain the mechanism behind the s-curve.}
\end{figure}

The equations are:
%\begin{equation}
\begin{equation}
\label{SDIR1}
\frac{d s_n}{dt} = -a(t)\,s_n i_n
\end{equation}
%\end{equation}
%\begin{equation}
\begin{equation}
\label{SDIR2}
\frac{d i_n}{dt} = a(t)\,s_n i_n - b\,i_n
\end{equation}
%\end{equation}
\begin{equation}
\label{SDIR3}
r_n = 1-s_n -i_n
\end{equation}
in normalized units, where $s_n$ is the susceptible fraction of the population, $i_n$ is the infected population and $r_n$ is the removed population (either by recovery or death), for each administrative division. $b$ is the recovery rate, while the infectious rate, $a(t)$, is defined as
\begin{equation}
\label{InfectionRate}
a(t) = a \,\{1 - [e^{-k(t-t')}-1] \theta(t-t')\},
\end{equation}
In Eq. \ref{InfectionRate}, $a$ represents the overall infection rate, while $k$ represents the rate at which control measures reduce transmission, $t'$ is a delay before which individuals are not aware enough of the disease to try to limit its spread, and $\theta$ is the Heaviside step function. Before $t'$, the disease infects at a rate $a$, while afterwards, we see a sharp drop-off in the disease transmissibility. It is clear that other, more realistic assumptions could be included into the model, including an exposed state (e.g., the SEIR model \cite{SEIR}, or the effect of burial practices, hospitalization, and other factors \cite{EbolaMixingModel1,EbolaMixingModel2,EbolaMixingModel3,EbolaMixingModel4}, but we leave this out of our model for simplicity. 
%With the recovery rate $\gamma$ kept constant, we model intervention through exponential fall-off of the infection rate $\beta$ with $k$, after some delay $t_0$:
We see rough quantitative agreement with empirical data (e.g., Fig.~\ref{fig:all-country-fits}), which suggests that this simple model captures the essence of the real-world dynamics.

In review, our model assumes only 3 states and a global time time-varying infection rate, $a(t)$. Even though there are reasons to consider this model too simplistic, it is able to effectively describe the roughly s-like curve of infections, and therefore it begins to help us understand the mechanism for a logistic-like cumulative distribution (Fig. \ref{fig:original-scaled-infection-curve-over-time}), such as reductions in infectability due to disease awareness.%, in agreement with previous work \cite{EbolaMixingModel1,EbolaMixingModel2,EbolaMixingModel3,EbolaMixingModel4}.

\subsection*{Are Strains Uniformly Transmissible?}

In this section, we use EVD genome sequences to determine what strains appear more often than expected by chance in Guinea and Sierra Leone. Our results suggest that EVD strains do not necessarily have uniform transmissibility. We are not aware of any previous models that take the strain of EVD into account, although a previous paper found certain EVD strains in Sierra Leone have different growth rates \cite{EbolaCompetingClaudes}. Unlike the previous paper, however, we can use our method to understand the transmissibility of strains in Guinea, where the sampling rate is otherwise too low.

\subsubsection*{Modeling Strain-Dependent Infection Probabilities}
We use meta data from Ebola nucleotide sequences isolated from patients in Guinea \cite{GuineaLineages1} between April, 2014 and January, 2015, and Sierra Leone between late May 2014 and January, 2015 \cite{SLEbolaSeq2014,SLEbolaFreeTown,SLEbolaSeq}, to determine when and where a strain of EVD was found, then use kernel-density estimation (KDE, see Methods) to estimate the spatial probability density function (PDF) of being infected with an EVD strain $s$\cite{KDE}:

\begin{equation}
\label{strainEVD}
P_{E}(\vec{x},t,h,\Delta t|s) = \frac{C}{|\mathcal{S}_s| h^2} \sum_{i\in \mathcal{S}_s} K\left(\frac{\vec{x}-\vec{x}_i}{h}\right) H\left(\Delta t - |t-t_i|\right),
\end{equation}
Here, $K$ represents a kernel with bandwidth $h$ to represent the area around an observed sequence where individuals are likely to be infected, $C$ is an overall constant and $H$ is the Heaviside step function. $\mathcal{S}_s$ are all labels for the set of pairs $\{\vec{x}_i,t_i\}$ with known infections of strain $s$. 
%is the total amount of data, regardless of the strain, between $\text{min}(\{t_i\})$ and $\text{max}(\{t_i\})$, because strains can mutate over time, limiting the usefulness of entire history of the infection. 
We add a Heaviside step function in the above equation to represent a sliding time window of length $\Delta t$ -- only within timeframe $(t - \Delta t)$ to $t$ is sequence data relevant. For the rest of the paper, our kernel is chosen to be a radially symmetric Gaussian:

\begin{equation}
K\left(\frac{\vec{x}-\vec{x}_i}{h}\right) = K\left(\frac{\|\vec{x} - \vec{x}_i\|}{h}\right) = \frac{1}{2 \pi}\text{Exp}\left(\frac{\|\vec{x} - \vec{x}_i\|^2}{2 h^2}\right).
\end{equation}

\begin{figure}[hbt]
\centering
\includegraphics[width=0.5 \columnwidth]{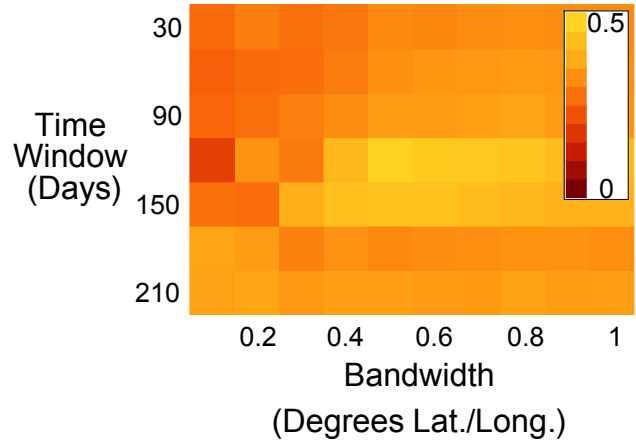}~~~~~~~~~~~~~~~~~~
\caption{\label{fig:GuineaProbabilityCorrel}
A heat map of the Spearman rank correlation between Eq.~\ref{Prob} and the number of individuals infected with EVD within a time $\Delta t$, as a function of the bandwidth $h$ and the time window width, $\Delta t$ in Eq.~\ref{Prob}. The correlation between the model and data is highest (Spearman $\rho = 0.50$, $p < 0.01$, $n = 68$) when the bandwidth is 0.5 degrees and the time window width is 120 days.}

\end{figure}

We do not believe that our results are qualitatively sensitive to the kernel choice. By summing these probabilities, we can then find the probability of being infected with EVD:

\begin{equation}
\label{Prob}
P_{E}(\vec{x},t,h,\Delta t) = \frac{C}{nh^2} \sum_s\sum_{i\in \mathcal{S}_s} K\left(\frac{\vec{x}-\vec{x}_i}{h}\right) H\left(\Delta t - |t-t_i|\right),
\end{equation}
where $n = \sum_s |\mathcal{S}_s|$. 
There are two free parameters: the kernel bandwidth $h$ and sliding time window width $\Delta t$, but knowing the true infection pattern allows for these parameters to be estimated (see Fig. \ref{fig:GuineaProbabilityCorrel} for EVD in Guinea). 

We apply KDE to EVD in Guinea and Sierra Leone, and test whether the estimated probabilities of becoming infected correlate with the number of infected individuals within each time window. A high correlation would represent close agreement between the KDE and actual spatial probability, and would increase our trust in these findings. In Guinea, we find Spearman correlations of up to 0.5 ($p<0.01$, $n = 68$, see Fig. \ref{fig:GuineaProbabilityCorrel}) which suggests that the KDE is in good agreement with data. In Sierra Leone, however, these correlations are negligable. The Guinea dataset will therefore be the focus of the rest of our paper.

From Fig. \ref{fig:GuineaProbabilityCorrel}, we find that the model best correlates with Guinea's infection data if $h=0.5$ and $\Delta t = 120$ days. %It would be appropriate to ask why the correlation peaks at these values?
% We do not believe a biological hypotheses can be tested against our data, but 
A plausible biological reason for the time window to be 120 days is that it may correspond to the effective viral shedding time, which can occur over timescale of roughly 100 days \cite{ViralShed}. For example, the virus has been found to spread via sexual contact, and has been seen in vaginal fluid for up to 33 days, and in semen for up to 82 days via cell culture, and 550 days via RNA. These values are extremes and might not be likely to be encountered often, however there has been at least one other report of RNA seen in semen 100 days afterwards \cite{EVDSTD}. The length scale of 0.5 degrees, however, is roughly the distance between prefectures, which probably explains why the peak is at this bandwidth. %Past this timescale, the viral shedding is likely negligible, therefore we do not need a time window much longer than this value. 
We also find that our results are robust to choices of $h$ and $\Delta t$, and we therefore let $h$ vary between 0.1 and 1 degrees and $\Delta t$ vary between 30 and 300 days.

\subsubsection*{Measuring the Relative Transmissibility of Strains}

We have just demonstrated the applicability of KDE to measuring the relative likelihood of being infected with EVD across districts. This is not in and of itself too interesting -- indeed other models may create better quantitative predictions of how EVD spreads \cite{Spatiotemporal}. Unlike previous work, however, this model can predict the relative transmissibility of individual EVD strains (Eq. \ref{strainEVD}). We will apply these predictions towards determining the relative transmissibility of EVD strains. When a strain $s$ appears much more than it should compared to other strains, then we can conclude that $s$ may be more transmissible.

\begin{figure}[hbt]
\centering
\includegraphics[width=0.75\columnwidth]{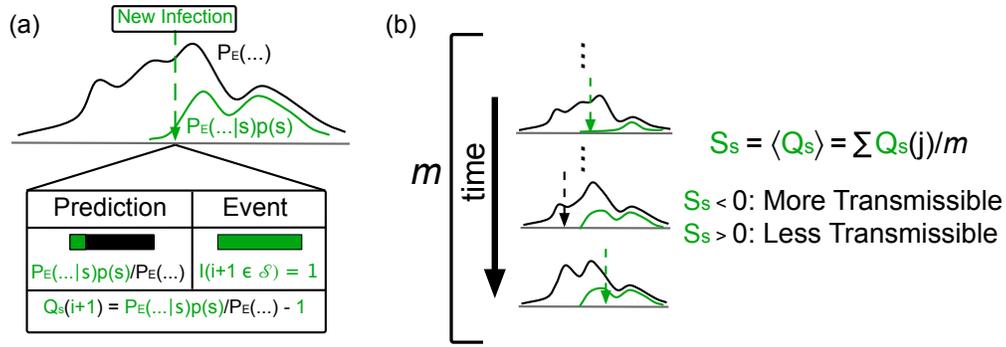}
\caption{\label{fig:EVDSchematic}A schematic for our success metric. (a) When a new infection appears, the probability the infection would be a strain $s$ is $P_{E}(\vec{x},t,h,\Delta t|s)$. We define $Q_s$ to be the difference between the prediction and actual event. (b) $S_s$, the expectation value of $Q_s$, should be 0 if $s$ is no more transmissible than the disease as a whole. If $S_s < 0$, then we would say that $s$ is more transmissible than the typical strain, while the opposite is true if $S_s > 0$.}
\end{figure}

Let a new infection arrive at position $\vec{y}$ and at time $t_y$. Our model may predict that the probability this infection would have a strain $s$ is, e.g., 20\% and any other strain is 40\%. If we find the disease is indeed strain $s$, we may be a little ``surprised". If $s$ consistently appears more often than expectations, we may believe that $s$ is simply more successful -- it can reach individuals more quickly than other strains. To quantify our small ``surprise" of seeing a new $s$ strain, we can write the equation:

%To measure the relative transmissibility of each strain, we calculate the relative probability an individual will be infected by strain A, versus any other strain. We do not compare between strains, because it may be difficult to know if a virus belongs to one strain or another, and furthermore, many infections were not found to belong to a major claude. If Strain A is found more (less) often than chance, we would say that it is stronger (weaker) than other strains overall. This is quantified by the equation:
\begin{equation}
\label{Qs}
Q_s(i+1) = \frac{P_E(\vec{x}_{i+1},t_{i+1},h,\Delta t|s) p(s)}{P_{E}(\vec{x_{i+1}},t_{i+1},h,\Delta t)} - I(i+1 \in \mathcal{S}_s),
\end{equation}
where $I$ is the indicator function that a new infection's strain is $s$ or not, and $p(s)$ is the probability that the infection strain is $s$ (See Fig. \ref{fig:EVDSchematic}). On average, the difference between our prediction and data, $S_s$, defined as

%In words, this quantity measure how often an infection is seen compared to chance.
%We define the success, $S$, of a strain as:

\begin{figure}[hbt]
\centering
\includegraphics[width=0.7\columnwidth]{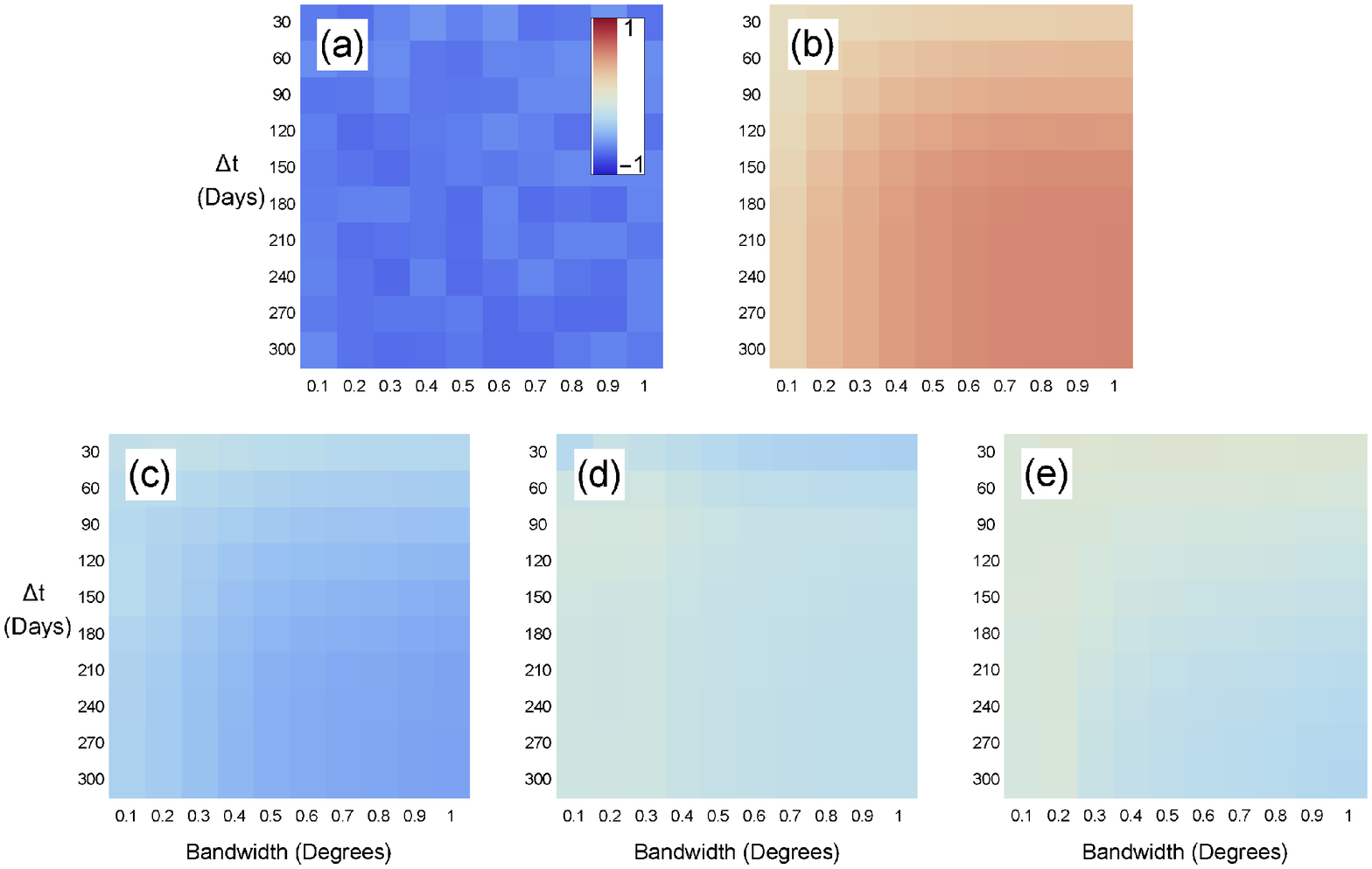}~~~~~~~~~~~~~~%GuineaStrainStrengths-SL1GN1234.eps}~~~~~~~~~~~~~~~~
\caption{\label{fig:GNPval}The success metric for strains in Guinea for (a) SL1, (b) GN1, (c) GN2, (d) GN3, (e) GN4, across various time window widths and bandwidths, where red values correspond to strains seen less often, and blue more often, than expected between the time the strain first appeared and last was seen. Sequence data comes from \cite{GuineaLineages1}.}
\end{figure}

\begin{equation}
S_s \equiv \langle Q_s\rangle = \frac{\sum_{j| t_{i_1}\le t_j\le t_{i_2}, i_1,i_2\in\mathcal{S}_s} Q_{s}(j)}{m},
\label{Success}
\end{equation}
should be 0, where the sum is over all sequences between the first and last appearance of strain $s$, and $m$ is the number of sequences that satisfy this constraint. When $S_s = 0$, $s$ is no stronger or weaker than the infection as a whole. If $S_s < 0$, then $s$ is stronger than strains of EVD generally, while $S_s>0$ implies the opposite (see Fig, \ref{fig:EVDSchematic}). To determine the statistical significance of $S_s$, we bootstrap values for which $S_s = 0$ (see Methods).
%but how big should the fluctuation be to be significant? 
%As explained in the methods section, we 
%where the sum is from the first time the strain is seen, until the last time, after $m$ sequences are sampled. 
%P-values can be determined by bootstrapping to find how often $S_s$ could be at least that large by chance. 
We find that $S_s$ is significantly different than chance, (Fig. \ref{fig:GNPval}), therefore, Ebola strains might not be uniformly transmissible.For example, SL1 has one of the most negative values of $S_s$ (Fig.~\ref{fig:GNPval}), where all values are statistically significant ($p<0.05$, $n$ varies with $\Delta t$). Interestingly, we also find the values of SL1 appear to fluctuate, possibly because we only have 9 SL1 datapoints, versus 37, 19, 13, and 36 datapoints for GN1, GN2, GN3, and GN4, respectively. Most other strains have a success metric indicating a stronger strain than expected by chance probably because few strains are concurrent in time, and therefore, comparing $S_s$ across strains is difficult. %This contrasts with the Sierra Leone data, where SL2 and SL3 were concurrent, and therefore presumably comparable.

 Our method may be compared to the method used by Meyer, Elias, and Hohle \cite{EpidPointProcess}, who compared the success of invasive meningococcal disease (IMD) strains. Common surveillance algorithms were found to be practical but these appear to be more useful for a chronic rather than an acute outbreak, such as EVD.

\section*{Discussion}

In conclusion, we find several factors not often accounted for that may improve the accuracy of modeling EVD. First, EVD appears to spread with a constant effective velocity through a migration network, in disagreement with the homogeneous mixing hypothesis (Figs. \ref{fig:DeffCountry} \& \ref{fig:DeffDistrict}, and Supplementary Fig. S1 online) commonly used to model diseases like Ebola. By taking into account the role a migration network has in disease spread, we can predict where EVD will arrive with greater accuracy than before. This method can also accelerate the process of identifying the index case by determining which administrative division is the origin through arrival time-effective distance correlation maximization. Second, we find that the growth of EVD at the finer spatial resolutions can be well described by three scaling parameters, and the initial growth rate decreases with the population density, contrary to our intuition, which suggests that a population density-dependent EVD model may more accurately predict the spread of Ebola when the disease first arrives at an administrative division. One plausible explanation for this result is that higher population density areas receive better healthcare than other areas, but more work is necessary to understand this behavior. Finally, we find a wide variation in the transmissibility of different strains of EVD, which suggests that modeling each disease strain, when this information is known, can improve the prediction task by reducing the heterogeneity of the data. In addition, our method may improve vaccination strategies if vaccines are made for particularly transmissible strains as well as the most common ones.

One way to take these factors into account would be through a meta-population model on top of a migration network with high spatial resolution. A meta-population model treats areas such as cities, districts, or countries as nodes on a network, with links connecting them to represent the flow of individuals from one area to another. Diseases within each area (node) are modeled with a compartmental model, under the assumption that the homogeneous mixing approximation is more accurate in smaller areas than for the entire network. Previous work has already found that a meta-population model \cite{ContagionGeometry}, or spatio-temporal model \cite{Spatiotemporal}, can accurately predict the spread of diseases. Finally, strain-dependent transmissibility may further improve model accuracy.

Our approach towards testing assumptions in disease modeling is not restricted to EVD, but can be applied toward other diseases of epidemiological concern. Future work is therefore necessary to test our methods on other diseases and check whether a meta-population model will better predict disease spread.

\section*{Methods}

 This section explains how data on the cumulative number of infected individuals at the first and second administrative level (e.g., counties or districts) was gathered, how the migration network was constructed, and how we found and used strain data in Fig. \ref{fig:GNPval}.

\subsection*{Infection Data from Humanitarian Data Exchange and World Health Organization}
%\paragraph{Data sources and coverage}.

 For the 2014-2016 EVD epidemic, we are aware of two main data sources on the cumulative number of infections: World Health Organization (WHO) patient database and the WHO weekly situation reports \cite{WHOEbola}. The patient database data produces results similar to the Situation Reports, but for consistency, all plots in this paper use the patient database. We focused on data from December 2013 to January 2016 for the major West African countries affected: Guinea, Liberia, Mali, Nigeria, Senegal, and Sierra Leone.

%\paragraph{Unique District Identifiers}. 

There was a significant amount of work parsing data from the patient database. Administration names in the data had multiple spellings, some administrative areas appeared individually but also aggregated with nearby areas, and finally spaces, accents, and other characters appeared inconsistently. These were cleaned and harmonized. Although in the most affected countries, we have the cumulative number of cases at the second administrative level, in three countries (Liberia, Mali, and Nigeria) we only have data at the first administrative level (region, not district). 

\subsection*{Migration Data from Flowminder}

%\textit{Data Source and Coverage}.

 Data on intra- and international migration in West Africa is taken from Flowminder \cite{EbolaMigration}. For modeling migration at fine spatial resolutions, the Flowminder data contain three different data sets: (1) within countries (capturing exclusively \underline{intra}national movement of people); (2) between countries (capturing exclusively \underline{inter}national movement of people); and (3) within and between countries, capturing both intra- and international movement of people. All three data sets include the West African countries most affected by EVD as well as  Benin, Cote d'Ivoire, Gambia, Ghana, Guinea Bissau, Senegal, and Togo, which had few, if any, EVD cases. In the third dataset, Flowminder collected census microdata from Public Use Microdata Series (IPUMS) on what country (or countries) an individual resided during the previous year.

To create a migration network, we first matched the Flowminder node coordinates to known district centroids (errors between centroids and node coordinates were $\pm 10$km). Having matched nodes to administrative names, we associated each node to the district-level arrival time of EVD, recording the date that the WHO patient database first records for a case in each administrative area. Four gravity model parameters were used to estimate the traffic between administration areas. Three were fit to migration using cell phone data in Cote d'Ivoire, Kenya, and Senegal, respectively, while the final one was fit to IPUMS data. We found that all produce similar fits, and work equally well to estimate the effective distances between areas. In addition, using population data from Geohive (next section), we created a radiation migration model, found to more accurately estimate migration patterns than gravity models \cite{MigrationModel}, we therefore used this for all figures that use migration networks in this paper. Figures using gravity models can be seen in Supplementary Figs. S1 \& S2  online.
 
\subsection*{Population data from GeoHive}
To determine the population density, and to estimate the migration network from the radiation migration model \cite{MigrationModel}, we first find the population and area of each administrative division. First, we collected the area administrative divisions in West Africa from www.geohive.com. Next, we collected population in those districts from population census records. For districts providing multiple census datasets, we collected from the latest report: Guinea, 2014; Liberia, 2008; Mali, 2009; Nigeria 2011; Senegal 2013; and Sierra Leone, 2004. Some population data is probably out-of-date and may lead to some biases in the population density and radiation model. However, we believe that newer data will confirm our initial conclusions.

\subsection*{Fitting Disease Models}

To find the best-fit parameters and associated errors in the logistic model (Eq. \ref{LogisticRegression}), and the SDIR model (Eqs. \ref{SDIR1}, \ref{SDIR2}, \& \ref{SDIR3}), we used least squares fitting. To reduce the possibility of overfitting data, we focus on districts where more than 20 individuals become infected.

\subsection*{Temporal-Spatial Resolution of Sequences}

To find empirical values of %the success of individual strains
Eqs. \ref{strainEVD} \& \ref{Prob}, we gathered infection meta-data (the strain, time, location of infected individual) from recent papers \cite{SLEbolaSeq2014,SLEbolaFreeTown,SLEbolaSeq,GuineaLineages1}. In the future we hope to make our own phylogenetic tree from the sequences, but for now we use the strain labels from the supplementary data itself.

A substantial fraction of sequences do not belong in any significant clade, and some sequences could belong to multiple clades, depending on the phylogenetic tree method used \cite{SLEbolaSeq2014,SLEbolaFreeTown,SLEbolaSeq}, therefore, we found $Q_s$ and $S_s$ for each strain by comparing that strain to all other sequences within the time window.

%Although the values seen in Fig. \ref{fig:GNPval} comes directly from Eq.~\ref{Success}, the p-values required a method that can tell us what the values of Eq.~\ref{Success} would typically be, if strains were not more likely to be seen compared to chance. To determine this
To determine the statistical significance of values in Fig. \ref{fig:GNPval}, we generate Bernoulli random variables (1 with probability $Pr(s,\vec{x}_i,t_i)$, and 0 with probability $1-Pr(s,\vec{x}_i,t_i)$) with
\begin{equation}
Pr(s,\vec{x}_i,t_i) = \frac{P_E(\vec{x}_{i},t_{i},h,\Delta t|s) p(s)}{P_{EVD}(\vec{x_{i}},t_{i},h,\Delta t)}
\end{equation}
 where  $\vec{x}_{i}$ is the infection location at a time $t_{i}$, to represent idealized data in which $\langle S_s\rangle=0$. %We then calculate $Q_s$, where the random variables represent ``idealized" data, in which each strain is no stronger than any other. 
We determine $S_{s,bootstrap}$ (Eq. \ref{Success}) from these idealized values of $Q_s$.% (which are not 0 due to random fluctuations). 
If the absolute value of the empirical $S_s$ value, $|S_{s,empirical}|$, is greater than 95\% of $|S_{s,bootstrap}|$ values, then the empirical data is statistically significant.
%, determine the success measure $S$, and record whether $S$ is greater than the empirical data. We bootstrapped using the $\{p_i\}$ values $10^4$ times to obtain our p-values.

%\bibliography{EbolaBib}

\section*{Acknowledgments}

The authors would like to thank the Santa Fe Institute and St. John's College for providing a productive working environment at the Complex System Summer School, and Samuel Scarpino for our many fruitful discussions there. The authors are also indebted to Caitlin Rivers' explanatory notes on the interpretation of the WHO patient database and situation reports. Finally, K. B. would like to thank Michelle Girvan for her many insightful suggestions.

\section*{Author contributions statement}

J.H., K.B. and C.V. conceived the models, C.V. and M.I. cleaned WHO Situation Reports and the patient database data on Ebola, all authors analyzed results, and all authors reviewed the manuscript.

\section*{Additional information}
\textbf{Competing financial interests} The authors declare no competing financial interests.

\newpage
\section*{Supplimentary Information}

\begin{figure}[h]
\centering
\includegraphics[width=0.8\columnwidth]{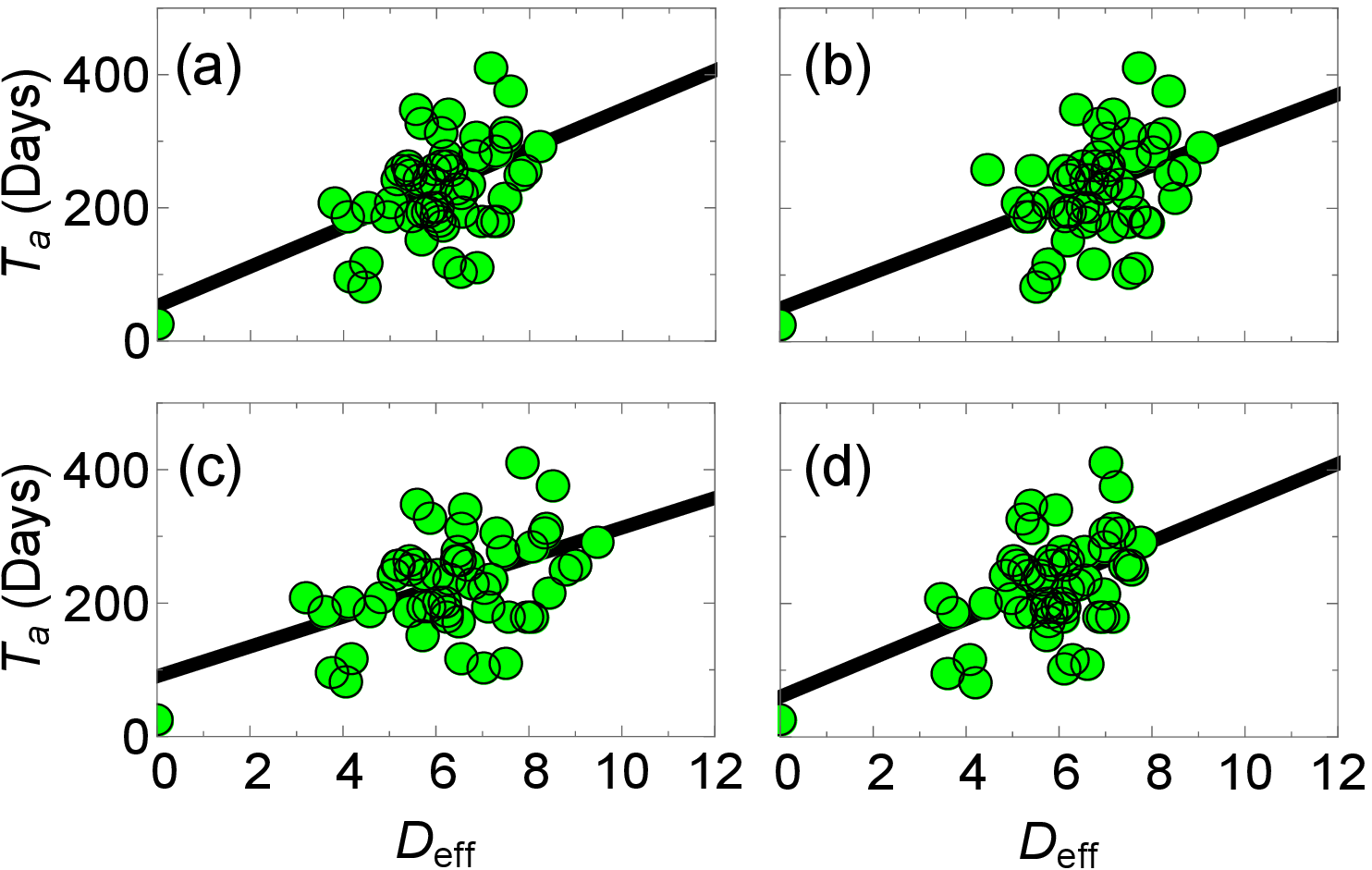}
%\caption{\label{fig:ArrivalTimeVsDeff}}
\end{figure}
\noindent
\textbf{Figure S1.} Arrival time (in days) versus the effective distance, using gravity migration models created in: Wesolowski, A. et al. Commentary: Containing the ebola outbreak -- the potential and challenge of mobile network data. PLoS Currents Outbreaks (2014). Migration models with parameters based on (a) Cote d'Ivoire, (b) IPUMS, (c) Kenya, and (d) Senegal datasets. Correlations are all statistically significant, with $p<10^{-4}$.
\newpage
\begin{figure}[h]
\centering
\includegraphics[width=0.8\columnwidth]{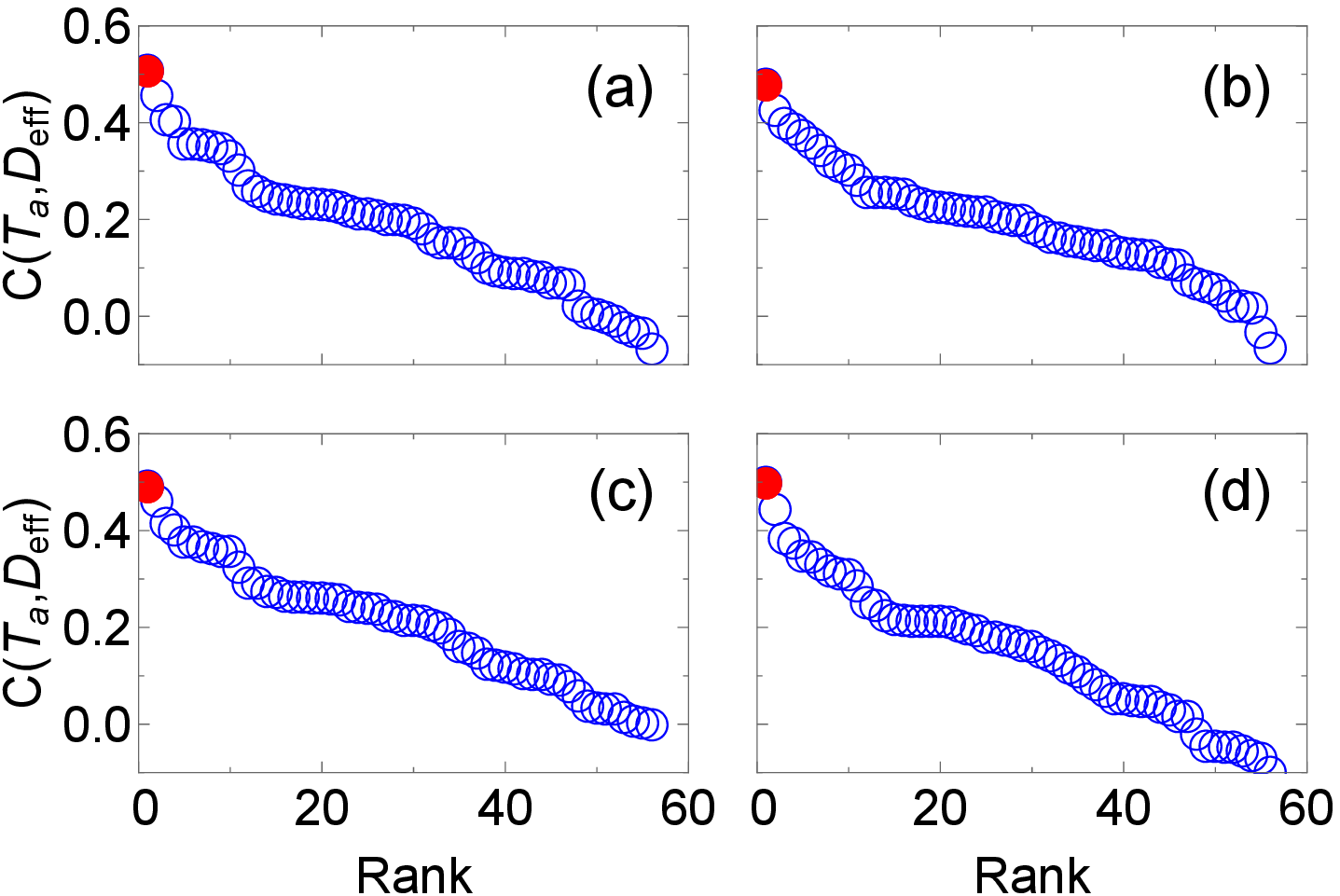}
\end{figure}
\noindent
\textbf{Figure S2.} The correlation between arrival time (in days) versus the effective distance, using migration models created in the previously cited paper, with various administrative divisions as candidate infection sources. In red is the true source of the infection, Gu\'eck\'edou, Guinea, which consistently has the highest correlation. Migration models shown here have parameters based on (a) Cote d'Ivoire, (b) IPUMS, (c) Kenya, and (d) Senegal datasets. Using the bootstrapping method described in our paper, we find the correct origin $68\%\pm1.5\%$, $63\% \pm 1.5\%$, $44\%\pm1.6\%$, and $70\% \pm 1.4\%$ of the time, respectively ($n=56$, where we bootstrap $10^3$ times). 

%\section*{SDIR Fits}
%Below are other fits for the SDIR model.
\newpage
\begin{figure}[h]
\centering
\includegraphics[width=0.8\columnwidth]{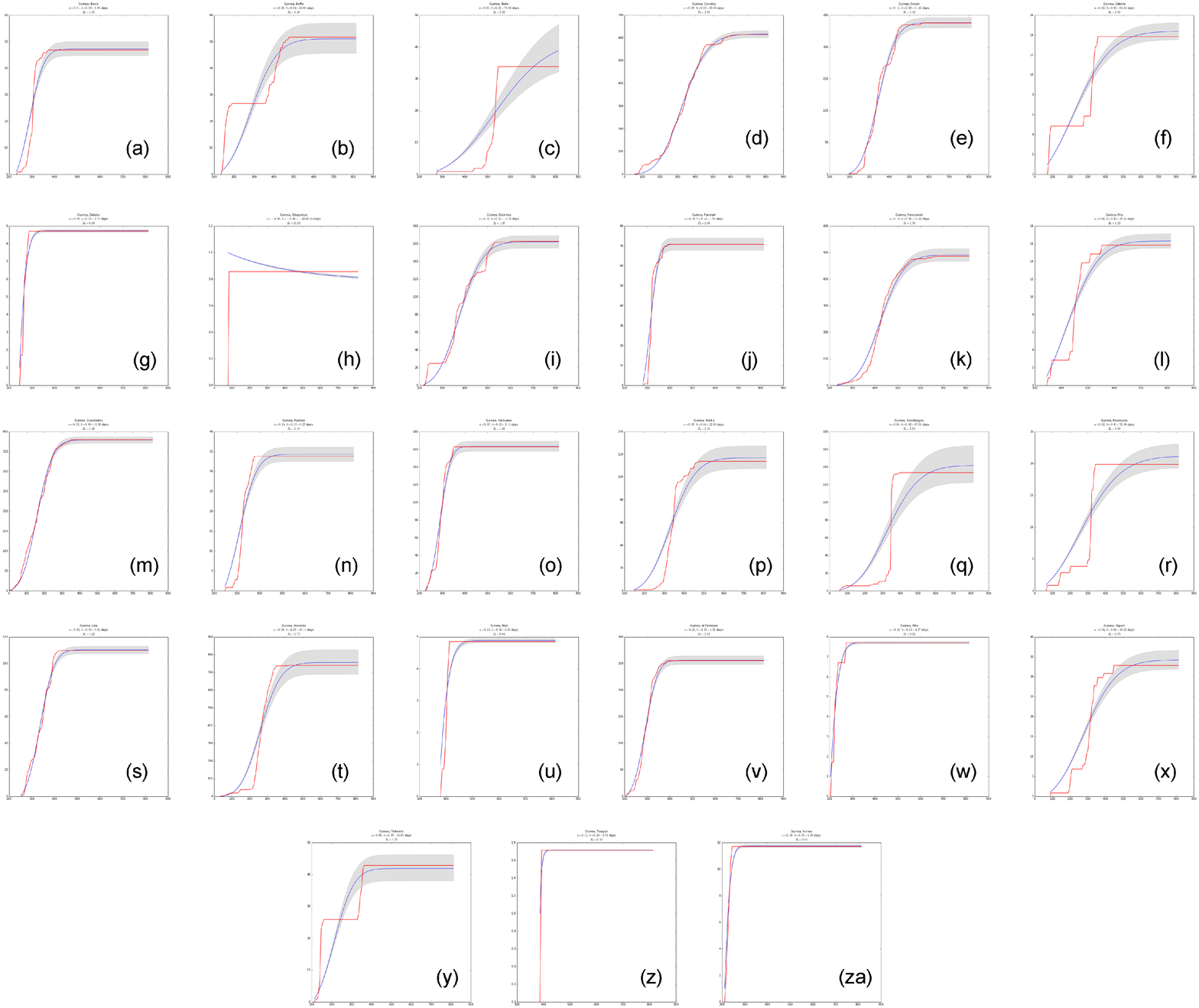}
\end{figure}
\noindent
\textbf{Figure S3.} The best fit for the SDIR model for all infected prefectures of Guinea. In order, these are: (a) Beyla,
 (b) Boffa,
 (c) Bok\'e,
 (d) Conakry,
 (e) Coyah,
 (f) Dabola,
 (g) Dalaba,
 (h) Dinguiraye,
 (i) Dubr\'eka,
 (j) Faranah,
 (k) For\'ecariah,
 (l) Fria,
 (m) Gu\'eck\'edou,
 (n) Kankan,
 (o) K\'erouan\'e,
 (p) Kindia,
 (q) Kissidougou,
 (r) Kouroussa,
 (s) Lola,
 (t) Macenta,
 (u) Mali,
 (v) Nz\'er\'ekor\'e,
 (w) Pita,
 (x) Siguiri,
 (y) T\'elim\'el\'e,
 (z) Tougu\'e, and 
 (za) Yomou.
\newpage
\begin{figure}[h]
\centering
\includegraphics[width=0.8\columnwidth]{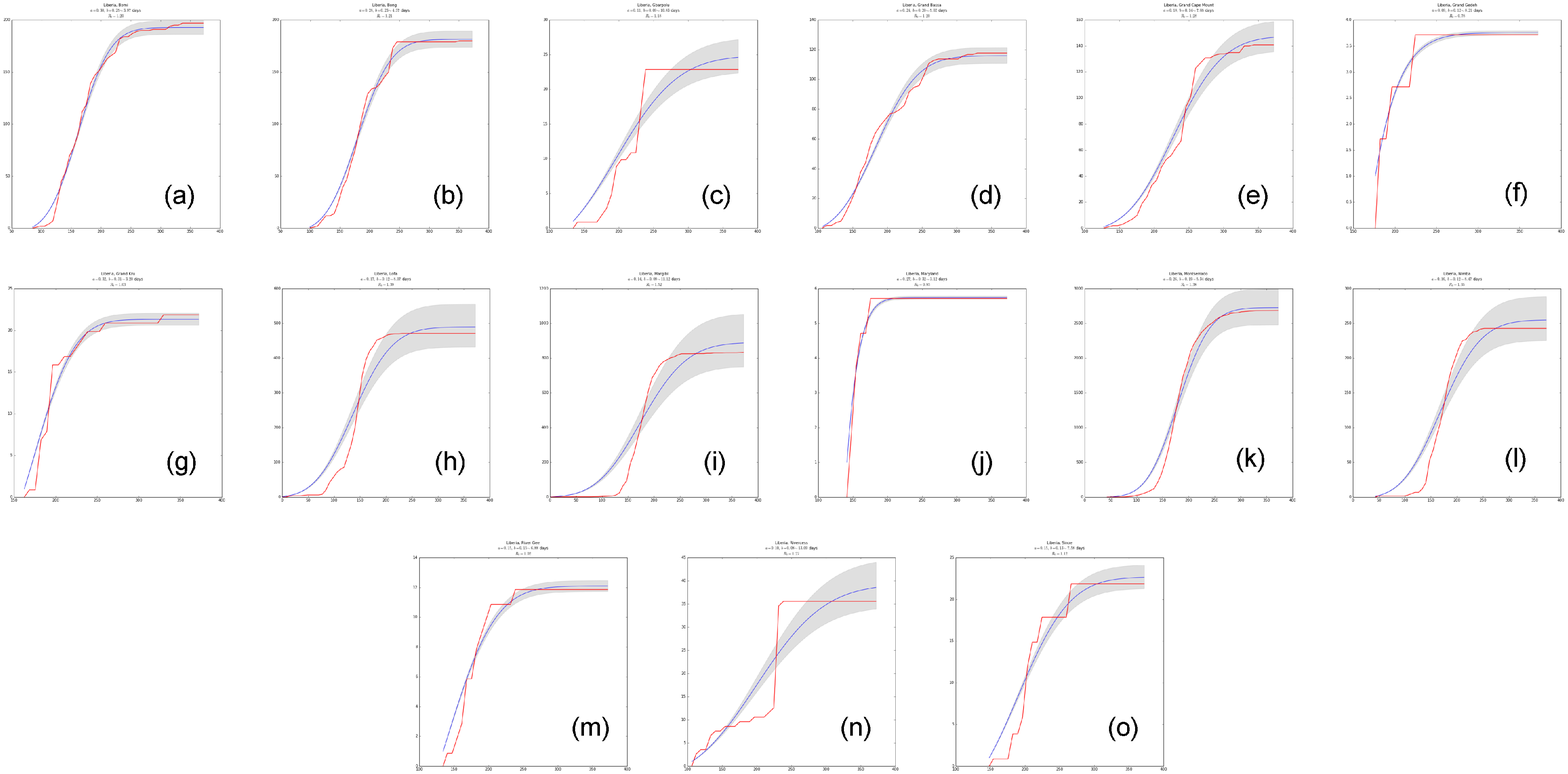}
\end{figure}
\noindent
\textbf{Figure S4.} The best fit for the SDIR model for all first administrative level divisions in Liberia. In order, these are: 
 (a) Bomi,
 (b) Bong,
 (c) Gbarpolu,
 (d) Grand Bassa,
 (e) Grand Cape Mount,
 (f) Grand Gedeh,
 (g) Grand Kru,
 (h) Lofa,
 (i) Margibi,
 (j) Maryland,
 (k) Montserrado,
 (l) Nimba,
 (m) Rivercess,
 (n) River Gee, and
 (o) Sinoe.
\newpage
\begin{figure}[h]
\centering
\includegraphics[width=0.8\columnwidth]{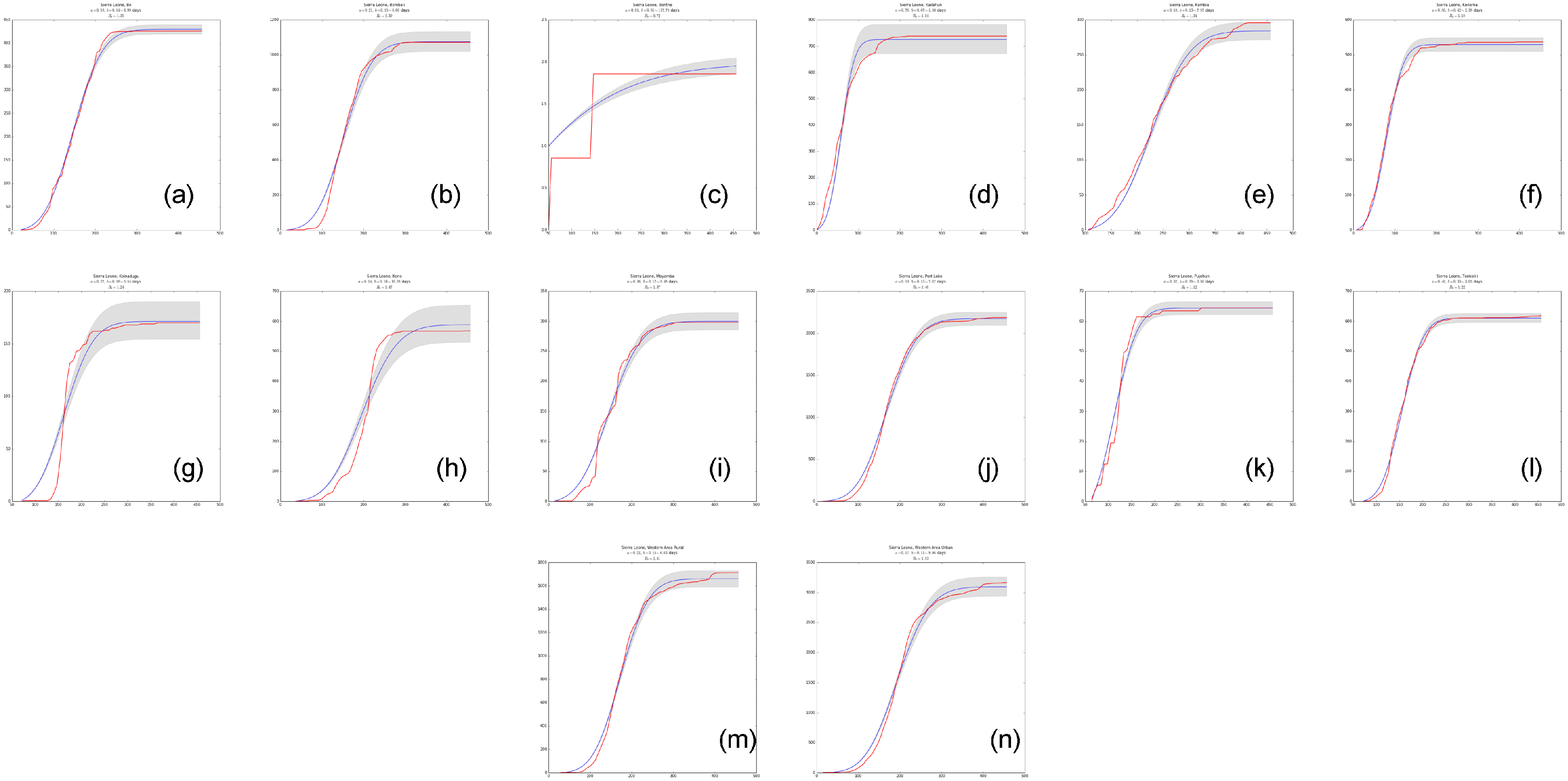}
\end{figure}
\noindent
\textbf{Figure S5.} The best fit for the SDIR model for all first administrative level divisions in Sierra Leone. In order, these are: 
 (a) Kailahun,
 (b) Kenema,
 (c) Kono	Eastern,
 (d) Bombali,
 (e) Kambia,
 (f) Koinadugu,
 (g) Port Loko,
 (h) Tonkolili,
 (i) Bo,
 (j) Bonthe,
 (k) Moyamba,
 (l) Pujehun,
 (m) Western Rural, and
 (n) Western Urban.

\end{document}